\documentclass[aps,prd,showkeys,amsmath,amssymb,groupedaddress]{revtex4}

\renewcommand{\section}[1]{\paragraph{#1. ---}\phantomsection\addcontentsline{toc}{section}{#1}}
\usepackage{ifthen}
\newboolean{prd}
\setboolean{prd}{false}
\newboolean{arxiv}
\setboolean{arxiv}{true}
\newboolean{notprd}
\setboolean{notprd}{true}
\ifprd
\setboolean{notprd}{false}
\fi
\newboolean{notarxiv}
\setboolean{notarxiv}{true}
\ifarxiv
\setboolean{notarxiv}{false}
\fi

\usepackage{graphicx}
\usepackage[usenames]{xcolor}
\usepackage{subfigure}
\usepackage{amssymb}

\usepackage[latin1]{inputenc}
\usepackage{amsfonts}
\usepackage{setspace}
\usepackage{amsmath}
\usepackage{rotating}
\usepackage{ulem}
\usepackage{color}
\xdefinecolor{mylinkcolor}{rgb}{0,0,0.5}
\usepackage[
        bookmarksnumbered, bookmarksopen, bookmarksopenlevel=2,
        breaklinks=true,
        colorlinks=true, filecolor=mylinkcolor, citecolor=mylinkcolor,
        linkcolor=mylinkcolor, urlcolor=mylinkcolor, menucolor=mylinkcolor,
]{hyperref}

\usepackage{wasysym}
\newcommand{\be}{\begin{equation}}
\newcommand{\ee}{\end{equation}}
\newcommand{\bea}{\begin{eqnarray}}
\newcommand{\eea}{\end{eqnarray}}

\newcommand{\prt}{\partial}

\allowdisplaybreaks

\begin{document}

\hypersetup{
        pdftitle={Accretion onto a spinning black hole},
        pdfauthor={Saha, Sen, Nag, Roychowdhury and Das}
}

\title{Model dependence of the multi-transonic behavior, stability properties and
corresponding acoustic geometry for accretion onto a spinning black hole.}

\author{Sonali Saha}
 \email{sonali[AT]sncwgs.ac.in}
\affiliation{Sarojini Naidu College for Women, Kolkata 700028, India} 

\author{Sharmistha Sen}
 \email{sharmistha1811[AT]gmail.com}
\affiliation{Camellia Institute of Technology, Kolkata 700129, India}

\author{Sankhasubhra Nag}
 \email{sankha[AT]sncwgs.ac.in}
\affiliation{Sarojini Naidu College for Women, Kolkata 700028, India}

\author{Suparna Roychowdhury}
 \email{suparna.roychowdhury[AT]gmail.com}
\affiliation{St. Xavier's College, Kolkata 700016, India}

\author{Tapas K Das} 
 \email{tapas[AT]hri.res.in {\sf [Corresponding Author]}}
\affiliation{Harish Chandra Research Institute, Allahabad 211019, India}

\date{\today}

\begin{abstract}
\noindent
Multi-transonic accretion for a spinning black hole has been compared among different disc geometries within post Newtonian pseudo potential framework. The variation of stationary shock characteristics with black hole spin has been studied in details for all the disc models and compared for adiabatic as well as isothermal scenario. The variations of surface gravity with spin for all these cases have also been investigated.
\end{abstract}

\keywords{Accretion disc, black hole physics, analogue gravity, stability}

\maketitle
\section{\bf Introduction}
\vskip 0.1cm
\noindent
The inner boundary condition imposed by the event horizon requires that accretion onto astrophysical black holes may manifest the transonic behavior in general \cite{liang-thomson,fkr02,kfm98}. For low angular momentum inviscid axisymmetric flow under the influence of the post-Newtonian pseudo-Schwarzschild black hole potentials multi-transonicity may be realized in the phase-portrait of the stationary accretion solutions- leading to the possibility of the formation of steady standing shock in between the saddle type outer sonic point and the center type middle critical point \cite{liang-thomson,Abramowicz1981,boz-pac,boz1,bmc86,blaes87,c89,ak89,abram-chak,das02,dpm03}. \\

The inviscid flow assumption, however, requires that the accretion flow is characterized by reasonably low angular momentum. Such angular momentum distribution should necessarily be sub-Keplerian. Also, such flow is expected to possess ab-initio non-zero value of the advective velocity along the equatorial plane of the flow. Such a configuration may perhaps be better described by a quasi-spherical flow rather than a standard thin disc structure with radius dependent flow thickness. A conical wedge shaped geometrical configuration may be the appropriate description for the axisymmetric  flow profile under consideration and such geometry may be a better alternative of the flow in hydrostatic equilibrium in the vertical direction (hereafter flow in vertical equilibrium in brief). Results obtained so far in the literature for inviscid axisymmetric black hole accretion in vertical equilibrium are, thus, to be compared with the results obtained using a more general theoretical set up capable of studying various features of the low angular momentum inviscid axisymmetric accretion in all three different flow geometries usually considered in the relevant literature, viz., accretion flow with constant thickness, conical wedge shaped flow, and accretion in vertical equilibrium (Ref \cite{nard12} and references therein). In most of the works dealing with the multi-transonic accretion flow deal with the stationary integral flow solutions. Transient phenomena are, however, not quite rare in large scale astrophysical set up. Hence, it is important to establish that the information obtained from the stationary solutions are reliable by ensuring that the steady states are stable for black hole accretion - at least for astrophysically relevant time scales. Such a task may be accomplished by perturbing the set of equations describing the steady state accretion flow and by demonstrating that such perturbations do not diverge with time. For the pseudo-Schwarzschild black hole accretion, it has recently been argued \cite{nard12} that the stationary inviscid flow described by a barotropic equation of state, are stable under linear perturbation of the mass accretion rate. Such observations remain valid for all three geometrical flow configurations described in previous paragraphs. Existence of certain mapping of the stationary transonic astrophysical flows (in strong gravity) with certain aspects of the analogue gravity (see, e.g., \cite{nvv02,su07,Barcelo2011} for a comprehensive discussion about the analogue gravity phenomena in general) has recently been established \cite{Bilic2014} for pseudo-Schwarzschild axisymmetric accretion onto a non-rotating black hole for different geometric profiles of the accreting matter. The corresponding acoustic geometry was realized by studying the transonic properties of the stationary accretion solutions, and the numerical value of the associated acoustic surface gravity $ k $ was estimated. Off late, the conjecture that most (if not all) of the astrophysical black holes possess non-zero values of the spin angular momentum ($ a \neq  0 $, where $ a $ is the Kerr parameter) has gained widespread acceptance \cite{mil09,kmtnm10,zio10,tch10,daly11,bggnps11,rey11,mcc11,mar11,dau10,nix11,tch12,
garofalo2013,McKinney-Tchekhovskoy-Blandford,Brenneman,Dotti-Colpi-Pallini-Perego-Volonteri,
Sesana-Barausse-Dotti-Rossi,Fabian-Parker-Wilkins-Miller-Kara-Reynolds-Dauser,
Healy-Lousto-Zlochower,Jiang-Bambi-Steiner, Nemmen-Tchekhovskoy}. It is thus essential to understand how the black hole spin influences the salient features of the multi-transonic stationary integral flow solutions as well as to check whether the linear perturbation scheme gets altered with the variation of $ a $. Which, in turn, would highlight how the space-time structure of the associated acoustic geometry depends on the rotation of the black hole. Complete general relativistic treatment to address such issues is indeed non-trivial. Although the stationary multi-transonic solutions are possible to obtain for flows in the Kerr metric \cite{dcz12}, full space-time dependent stability analysis of the transonic solutions does not seem to be amenable within the framework of the general relativity, especially when attempts are made to provide a complete analytical solution. \\

Motivated by the aforementioned arguments, in the present work we study low angular momentum inviscid axisymmetric hydrodynamic accretion onto a rotating black hole using post-Newtonian pseudo-Kerr potentials.  Several such potentials are available in the literature \citep{Chakrabarti1992,abn96,Semerak1999,Lovas1998,Mukhopadhyay2002,Chakrabarti2006,Ghosh2007,Ghosh2014,Karas2014}. Among all potentials proposed, we however, pick up the one introduced by Artemova et. al \cite{abn96} since it has the algebraically simplest form. The primary aim behind the introduction of a pseudo-Schwarzschild / Kerr black hole potential is to compromise between the convenience of handing of a Newtonian description of gravity and the space time structure described by complicated general relativistic calculations. Introduction of such potentials should allow one to investigate the basic physical process which occur during the accretion phenomena within the framework of a Newtonian construct, hereby avoiding the formal general relativistic description of the dynamics of the inflowing matter. The intrinsic simplicity of the algebraic presentation of a pseudo-potential is of paramount importance to let that potential qualify as the useful one. No particular pseudo-potential stands alone as a significantly superior one (compared to the rest of such potentials) to mimic all the properties of the full general relativistic Kerr metric - especially while describing the dynamics of the multi-transonic flow - the potential with the simplest algebraic form will be of significant interest for the aforementioned purpose. \\

The viscous transport of angular momentum is not taken into account in this paper. Forty one years after the introduction of the Standard Disc model \cite{ss73,nt73}, exact modelling of viscous transonic accretion flow incorporating proper heating and cooling effects is still an arduous task. Also, as will be demonstrated later in this work, construction of acoustic geometry becomes practically impossible for dissipative flow since the viscous effect may destroy the Lorenz in variance. Large radial (advective) velocity close to the horizon ensures that the viscous time  is larger than the dynamical infall time scale. Large radial velocity even at the larger distance is due to the fact that the rotational energy of the axisymmetric flow remains relatively low \cite{bi91,ib97,pb03}. Aforementioned arguments indicate that the assumption of inviscid flow is not unjustified for low angular momentum flow with initial non-vanishing advective velocity profile. \\ 

For large scale relativistic astrophysical flows, however, the effect of the viscosity and magnetic field may not always be neglected. As a consequence, dissipative mechanism (through comptonization, bremsstrahlung or synchrotron processes) may become significant to influence the overall flow dynamics and hence the flow may be subjected to the turbulent instabilities. Non-linear perturbation may also be of considerable importance to determine the stability criteria of the flow. Velikov-Chandrasekhar \cite{Chandrasekhar1960} and / or  Balbus - Hawley \cite{Balbus1998} type magneto-rotational instabilities may be observed close to the black hole event horizon. These additional complexities clearly bring the system to the risk of complete destruction of the Lorenz symmetries and hence the analytical modelling of the flow profile as well as of the sonic geometry embedded into it becomes impossible - one has to take recourse to the large scale numerical simulation to handle such flow profile, which, clearly, is beyond the scope of this work. Also to mention in this aspect that the axisymmetric configuration of the flow is assumed a-priori and no misaligned flow structure may be handled using the formalism presented in this paper. \\

In what follows, we write down the Euler and the continuity equation for matter inflow considered in this work. Both polytropic and isothermal flow will be analyzed for three different geometric configurations of the flow. These two equations are then integrated to obtain the corresponding first integrals of motion for the respective flow configurations. The critical point conditions are derived and the parameter space (spanned by various initial boundary conditions governing the flow) for the multi-transonic accretion are identified. The regions of the parameter space for which a steady standing stock may form are determined. A linear eigenvalue scheme to identify the nature (whether they are of saddle or of center type) of the critical points is developed. Such a scheme helps to understand the nature of the two dimensional Mach number vs. radial distance phase portrait from an algebraic point of view - without explicitly evaluating the integral solutions of the governing flow equations. Formation of stationary shock is discussed and the parameter sub-space corresponding to the shocked flow is identified. Such a parameter sub-space for various flow configurations are compared to understand how the geometric configuration of accreting matter influences the shock formation mechanism. The spin dependence of the salient features of the multi-transonic shocked flow is investigated. \\

Once the properties of the stationary solutions are fully understood, a time dependent linear perturbation analysis of the stationary integral flow solutions is performed for a generic form of the pseudo-Kerr black hole potential to check whether such perturbations remain non-divergent, to estimate the stability of the steady state accretion solutions. It is further demonstrated that a space time dependent acoustic metric can be constructed to describe the propagation of the aforementioned linear perturbation and the corresponding pseudo-Riemannian sonic manifold is studied. Finally, the black hole spin dependence of the corresponding acoustic surface gravity is studied in somewhat detail. \\

Hereafter, the radial distance (measured along the equatorial plane of the flow) and any characteristic velocity will be scaled in units of  $ \dfrac{2 G M_{BH}}{c^2} $ and by $ c $, respectively. Where $ G $, $ M_{BH} $ and $ c $ are the universal gravitational constant, mass of the black hole considered, and the velocity of light in vacuum, respectively. We adopt natural unit by setting $ G = c = 1 $. \\ 

The expression for the free fall acceleration as provided by Artemova et-al \cite{abn96} is

\begin{equation}
\label{accelerationartemova}
f = - \dfrac{1}{r^{2 - \beta} (r - r_+)^\beta}
\end{equation}

The potential corresponding to the above acceleration diverges at the event horizon.

In the above expression

\begin{equation}
r_+ = 1 + \sqrt{1 + a^2}
\end{equation}

\begin{equation}
\beta = \dfrac{r_{in}}{r_+} - 1
\end{equation}

\begin{equation}
r_{in} = 3 + Z_2 - [(3 - Z_1)(3 + Z_1 + 2Z_2)]^{1/2}
\end{equation}

\begin{equation}
Z_1 = 1 + (1 - a^2)^{1/3}[(1 + a)^{1/3} + (1-a)^{1/3}]
\end{equation}

\begin{equation}
Z_2 = (3a^2 + Z_1^2)^{1/2}
\end{equation}

Explicit expression for the associated black hole potentials is given as

\begin{equation}
\label{phi}
\phi = \frac{r^{\beta - 2}\left[\frac{r^2}{r_+ - \beta r_+} - \frac{r_+r}{r_+ - \beta r_+}\right]}{(r -
r_+)^\beta} - \frac{1}{2}
\end{equation}

The Euler and the continuity equation for a generalized pseudo-Kerr potential $ \phi $ can be expressed as

\begin{equation}
\label{euler}
u\frac{du}{dr} + \frac{1}{\rho} \frac{dP}{dr} + \frac{d\phi(r)}{dr} = 0
\end{equation}

\begin{equation}
\label{equation of continuity}
\frac{d[u\rho rh(r)]}{dr} = 0
\end{equation}

In the present paper, the polytropic as well as the isothermal accretion will be considered. The equation of state corresponding to a polytropic flow is $ p = K \rho^\gamma $, $ p $, $ \rho $ and $ \gamma $ being the pressure, mass, density and the adiabatic index ( $ \gamma = C_p/C_v $, $ C_p $ and $ C_v $ are the specific heats at constant pressure and volume, respectively), respectively. The proportionality constant $ K $ may be interpreted as a measure of the entropy content of the flow since entropy per particle of an ensemble may be expressed as \cite{landauPK},

\begin{equation}
\label{entropyperparticle}
\sigma = \dfrac{1}{\gamma - 1} \log K + \dfrac{\gamma}{\gamma - 1} + \text{constant}
\end{equation}

Where the value of the constant depends on the chemical composition of the matter. \\

For the isothermal flow, the equation of state is expressed as 

\begin{equation}
\label{eos}
p = \frac{\rho RT}{\mu} = C_s^2 \rho
\end{equation}

Where $ C_s $ is the position independent (since $ C_s \propto T^{1/2} $) isothermal sound speed. $ R $ and $ \mu $ are the universal gas constant and the mean molecular weight, respectively. $ T $ is the proton temperature. \\

One needs to solve the time independent part of the above two equations to obtain the stationary integral flow solution (which is necessary to construct the corresponding phase portrait for the transonic accretion). The time independent Euler and the continuity equations are examples of the homogeneous differential equation of first order in flow velocity. Integral solutions of such differential equations usually provide the corresponding first integrals of motion, - i.e., the quantities which remain conserved. \\

For axisymmetric inviscid hydrodynamic accretion we thus have two first integrals of motion obtained by integrating the Euler and the continuity equations. \\

Since the continuity equation implies the conservation of mass, the corresponding first integral of motion - irrespective of the equation of state is used to describe the flow. The integral solution of the Euler equation will provide the total specific energy as another first integral of motion for a polytropic flow. \\

For isothermal accretion, the corresponding first integral of motion is not to be identified with the specific energy since energy exchange is allowed for such flow to maintain the space invariance of the bulk temperature of the flow. \\

The mass accretion rate, as we will observe, will be a function of the flow thickness, and will resume different values for three different geometric configurations of accretion.  On the other hand, the first integral of motion obtained from the Euler equation does not depend on the flow geometry - it rather depends on the equation of state (though the integration of the term $ \int \dfrac{dp}{\rho} $ contained in the Euler equation).

\section{\bf Adiabatic Flow}
The specific energy first integral of motion can be obtained as

\begin{equation}
\label{eadiabatic}
\mathcal{E} =  \dfrac{u^2}{2} + \dfrac{C_s^2}{\gamma - 1} + \frac{\lambda^2}{2r^2} + \phi(r)
\end{equation}
where $ u $ is the advective velocity defined on the equatorial plane.

The mass accretion rates for  three different flow geometries are obtained as

\begin{equation}
\label{MdotCH}
\dot{M}_{CH} = Hr\rho u
\end{equation}

\begin{equation}
\label{MdotCO}
\dot{M}_{CO} = \Theta \rho ur^2
\end{equation}

\begin{equation}
\label{MdotVE}
\dot{M}_{VE} = \sqrt{\dfrac{1}{\gamma}} u C_s \rho^{3/2} \phi'^{-1/2}
\end{equation}

where $ H $ is the half-thickness of the flow and is assumed to be a constant, $ \Theta = H(r)/r $ , where $ H(r) $ is the local flow thickness. 
The corresponding entropy accretion rates can be obtained as
\begin{equation}
\label{entropyaccretion}
\mathcal{\dot{M}} = \dot{M}_{\text{in}} \gamma^{\dfrac{1}{\gamma-1}} K^{\dfrac{1}{\gamma-1}} 
\end{equation}
\footnote[1]{ $ K $ is a measure of the entropy per particle as expressed through \ref{entropyaccretion}. Any function of $ K $, when multiplied by the total amount of mass flowing in per unit time, provides a measure of the total amount of inward entropy flux per unit time. $ \mathcal{\dot{M}} $ is thus called the entropy accretion rate. The concept of the entropy accretion rate was first introduced by Blaese \cite{blaes87}} 
The subscript $ [CH,CO, VE] $ implies that quantities are evaluated for flow with constant height, in conical equilibrium and in vertical equilibrium, respectively. \\

The relation between the space gradient of the dynamical velocity and that of the sound speed may be obtained by differentiating the expression for the corresponding entropy accretion rates for three different flow geometries. 

\begin{equation}
\left(\frac{dC_s}{dr}\right)_{CH} = \left(1-\gamma\right)\frac{C_{s}}{u}\left(\frac{1}{2}\frac{du}{dr}+\frac{u}{2r}\right)
\end{equation}

\begin{equation}
\left(\frac{dC_s}{dr}\right)_{CO} = \left(1-\gamma\right)\frac{C_{s}}{u}\left(\frac{1}{2}\frac{du}{dr}+\frac{u}{r}\right)
\end{equation}

\begin{equation}
\left(\frac{dC_s}{dr}\right)_{VE} = \left(\frac{1-\gamma}{1+\gamma}\right)\frac{C_{s}}{u}\left[\frac{du}{dr}+\frac{u}{2}\left(\frac{3}{r}-\frac{\phi"(r)}{\phi'(r)}\right)\right]
\end{equation}

We obtain the velocity space gradient (for three different flow geometries) by differentiating the algebraic expression for $ \mathcal{E} $ and by substituting the corresponding values of $ \dfrac{dC_S}{dr} $ (as obtained by differentiating the respective expressions for $ \dot{\mathcal{M}} $)

\begin{equation}
\label{du/dr CH Adiabatic}
\left(\frac{du}{dr}\right)_{CH} = \frac{\frac{\lambda^2}{r^3} + \frac{C_s^2}{r} - \phi'(r)}{u -
\frac{C_s^2}{u}}
\end{equation}

\begin{equation}
\label{du/dr CO Adiabatic}
\left(\frac{du}{dr}\right)_{CO} = \frac{\frac{\lambda^2}{r^3} + \frac{2C_s^2}{r} - \phi'(r)}{u -
\frac{C_s^2}{u}}
\end{equation}

\begin{equation}
\label{du/dr VE Adiabatic}
\left(\frac{du}{dr}\right)_{VE} = \frac{\frac{\lambda^2}{r^3} + \phi' -\frac{C_s^2}{\gamma + 1}
\left[\frac{3}{r} + 
\frac{\phi"}{\phi'}\right]}{u - \frac{2C_s^2}{u(\gamma +1)}}
\end{equation}

Following the usual procedure adopted in the literature (see e.g. \cite{nard12,crd06}) the critical point conditions can be obtained as 

Constant Height Model:
\begin{equation}
\label{u at critical point CH Adiabatic}
u_c^2 = C_{sc}^2 = r_{c} \phi'(r_c) - \frac{\lambda^2}{r_c^2}
\end{equation}

Conical Model:
\begin{equation}
\label{u at rc CO Adiabatic}
C_{sc}^2 = u_c^2 = \frac{1}{2}\left[r_c \phi'(r_c) - \frac{\lambda^2}{r_c^2}\right]
\end{equation}

Vertical Equilibrium Model:
\begin{equation}
\label{u at rc VE Adiabatic}
\sqrt{\frac{2}{1 + \gamma}} (C_s)_c = u_c = \left(\frac{\phi'(r_c) + \gamma \phi'(r_c)}{r_c^2}\right)
\left(\frac{\lambda^2 + r_c^3 \phi'(r_c)}{3\phi'(r_c) + r_c\phi"(r_c)}\right)
\end{equation}

Substitution of the critical point conditions into the expression for $ \mathcal{E} $ will provide an algebraic expression of the following generalized form

\begin{equation}
\mathcal{E} - f(r_c, \lambda, \gamma) = 0
\end{equation}

The explicit form of such equations for three different flow models under consideration may be obtained as \\

Constant Height Model:

\begin{equation}
\mathcal{E} - \label{polynomial for rc CH Adiabatic}
\frac{1}{2} \frac{\gamma+1}{\gamma-1}\left[r_{c,CH} \phi'(r_c,CH) - \frac{\lambda^2}{r_{c,CH}^2}\right] + 
\phi(r_{c,CH})
+ \frac{\lambda^2}{2r_{c,CH}^2} = 0
\end{equation}

Conical Model:

\begin{equation}
\mathcal{E} - \label{polynomial for rc CO Adiabatic}
\frac{1}{4} \left(\frac{\gamma + 1}{\gamma - 1}\right) \left[r_{c,CO} \phi'(r_{c,CO}) - \frac{\lambda^2}
{r_{c,CO}^2}
\right] + \phi(r_{c,CO}) + \frac{\lambda^2}{2r_{c,CO}^2} = 0
\end{equation}

Vertical Equilibrium Model:

\begin{equation}
\label{polynomial for rc VE Adiabatic}
\mathcal{E} - \left[\frac{\lambda^2}{2r_{c,VE}^2} + \phi(r_{c,VE}) \right] - \frac{2\gamma}{\gamma^2 - 1}
\left[\frac{\phi'(r_{c,VE}) 
+ \gamma \phi'(r_{c,VE})}{r_{c,VE}^2}\frac{\lambda^2 + r_{c,VE}^3 \phi'(r_{c,VE})}{3\phi'(r_{c,VE}) + 
r_{c,VE}\phi"(r_{c,VE})}\right] = 0
\end{equation}

A set of values of $ [\mathcal{E}, \lambda, \gamma, a] $ is required to solve the above algebraic expressions to obtain the value of the corresponding critical point $r_c$. 

${\cal E}$ is scaled by the rest mass energy and includes the rest mass
energy itself, hence ${\cal E}=1$ corresponds to a flow with zero
thermal energy at infinity, which is obviously not a realistic initial
boundary condition to generate the acoustic perturbation. Similarly,
${\cal E} <1$ is also not quite a good choice since such configuration
with the negative energy accretion state requires a mechanism for
dissipative extraction of energy to obtain a positive energy
solution\footnote{A positive Bernoulli's constant flow is essential
to study the accretion phenomena so that it can incorporate the
accretion driven outflows (see \cite{das99} and references therein).}.
Presence of any such dissipative mechanism is not desirable to study the
inviscid flow model considered in the present work.
On the other hand, almost all ${\cal E}>1$
solutions are theoretically allowed. However, large values of
${\cal E}$ represents accretion with unrealistically hot flows in
astrophysics. In particular, ${\cal E}>2$ corresponds to extremely
large initial thermal energy which is not quite commonly observed in
accreting black hole candidates. We thus set $1{\lesssim}{\cal E}{\lesssim}2$.

A somewhat intuitively obvious range for $\lambda$ for our
purpose is $0<\lambda\;{\le} 4$,
since $\lambda=0$ indicates spherically symmetric flow and for $\lambda>4$
the sub-Keplerian nature is lost and
multi-critical behavior does not show up in general.

$\gamma=1$ corresponds to isothermal accretion where the acoustic perturbation
propagates with position independent speed. $\gamma<1$ is not a
realistic choice in accretion astrophysics. $\gamma>2$ corresponds to
the super-dense matter with considerably large magnetic field and a
direction dependent anisotropic pressure. The presence of a dynamically
important magnetic field requires the solution of general relativistic
magneto hydrodynamics equations which is beyond the scope of the present work.
Hence a choice for $1{\lesssim}\gamma{\lesssim}2$ seems to
be appropriate. However, preferred bound for realistic black hole
accretion is from $\gamma=4/3$ (ultra-relativistic flow) to $\gamma=5/3$
(purely non relativistic flow), see, e.g., \cite{fkr02}
for further detail. Thus we mainly concentrate on $4/3{\le}\gamma{\le}5/3$.

The domain for $a$ lies clearly in between the values of the Kerr
parameters corresponding to the maximally rotating black hole for
the prograde and the retrograde flow. Hence the obvious choice
for $a$ is $-1{\le}a{\le}1$.
Although to be mentioned here that an upper limit for the Kerr
parameter has been set to $0.998$ in some works, see, e.g.,
\cite{thorne74}. We, in our work, however, do not consider any such
interaction of accreting material with the black hole itself which
might allow the evolution of the mass and the spin of the hole
as was considered in \cite{thorne74} to arrive at the conclusion
about such upper limit for the black hole spin. 

The allowed domains for the four parameter initial boundary conditions are thus 
$\left[1{\lesssim}{\cal E}{\lesssim}2, 0<\lambda{\le}4,\right.,$  $ \left. 4/3{\le}\gamma{\le}5/3,-1{\le}a{\le}1\right]$.

The four parameter set 
$\left[{\cal E},\lambda,\gamma,a\right]$
may further be classified
intro three different categories, according to the way they influence the
characteristic properties of the stationary transonic solutions.
$\left[{\cal E},\lambda,\gamma\right]$ characterizes
the flow, and not the spacetime since the accretion is assumed to
be non-self-gravitating.
The Kerr parameter $a$ exclusively
determines the nature of the spacetime and hence can be thought of
as some sort of `inner boundary condition' in a qualitative sense since 
the effect of
gravity truly requires the full general relativistic
framework only out to several gravitational radii, beyond which
it asymptotically approaches the Newtonian description.
$\left[{\cal E}, \lambda\right]{\subset}\left[{\cal E}, \lambda,\gamma\right]$
determines the dynamical aspects of the
flow, whereas $\gamma$ determines the thermodynamic properties.
To follow a holistic approach, one needs to study the variation of the
relevant features of the transonic accretion on all of these four parameters.

Once the values of the critical points are obtained (physically acceptable values of $ r_c $ should be validated by the criteria $ r_c > r_+ $), transonic accretion solutions can be obtained by integrating the corresponding expressions for $ \dfrac{du}{dr} $ subjected to the critical value of the $ \dfrac{du}{dr} $ , i.e., the value of $ \dfrac{du}{dr} $ evaluated at the critical point(s). \\

Expressions for the critical values of the space gradient of the dynamical velocities can be obtained as (see \cite{Bilic2014}) \\
 \begin{eqnarray}\label{dudrc-adiabatic-constant-height}
\left|\left(\frac{du}{dr}\right)_{\rm CH}\right|_{r_c}&=&\frac{1}{r_c}\left(\frac{1-\gamma}{1+\gamma}\right)\sqrt{{r_c\Phi'(r_c)}-\frac{\lambda^2}{r_c^2}}\\ \nonumber &&\pm\sqrt{\frac{1}{r_c^2}\left(\frac{1-\gamma}{1+\gamma}\right)^2\left({r_c\Phi'(r_c)}-\frac{\lambda^2}{r_c^2}\right)-\frac{\left(\frac{\gamma}{r_c^2}(r_c\Phi'(r_c)-\frac{\lambda^2}{r_c^2})+\frac{3\lambda^2}{r_c^4}+\Phi''(r_c)\right)}{\sqrt{{r_c\Phi'(r_c)}-\frac{\lambda^2}{r_c^2}}}} , 
\end{eqnarray}
\begin{eqnarray}\label{dudrc-adiabatic-conical-model}
\left|\left(\frac{du}{dr}\right)_{\rm CM}\right|_{r_c}&=&\frac{2}{r_c}\left(\frac{1-\gamma}{1+\gamma}\right)\sqrt{\frac{r_c\Phi'(r_c)}{2}-\frac{\lambda^2}{2r_c^2}}\\ \nonumber &&\pm\sqrt{\frac{4}{r_c^2}\left(\frac{1-\gamma}{1+\gamma}\right)^2\left(\frac{r_c\Phi'(r_c)}{2}-\frac{\lambda^2}{2r_c^2}\right)-\frac{\left(\frac{2(2\gamma-1)}{r_c^2}(\frac{r_c\Phi'(r_c)}{2}-\frac{\lambda^2}{2r_c^2})+\frac{3\lambda^2}{r_c^4}+\Phi''(r_c)\right)}{(1+\gamma)\sqrt{\frac{r_c\Phi'(r_c)}{2}-\frac{\lambda^2}{2r_c^2}}}} , 
\end{eqnarray}
\begin{eqnarray}\label{dudrc-adiabatic-vertical-equilibrium}
\left|\left(\frac{du}{dr}\right)_{\rm VE}\right|_{r_c}&=&2u_c\left(\frac{\gamma-1}{8\gamma}\right)\left[\frac{3}{r_c}+\frac{\Phi'''(r_c)}{\Phi'(r_c)}\right]\\ \nonumber &\pm& \sqrt{\frac{\gamma+1}{4\gamma}}\left[u_c^2\frac{\gamma-1}{\gamma+1}\frac{\gamma-1}{4\gamma}\left(\frac{3}{r_c}+\frac{\Phi''(r_c)}{\Phi'(r_c)}\right)^2\right.  \\ \nonumber  &-& \left. u_c^2\frac{1+\gamma}{2}\left(\frac{\Phi'''(r_c)}{\Phi'(r_c)}-\frac{2\gamma}{(1+\gamma)^2}\left(\frac{\Phi'''(r_c)}{\Phi'(r_c)}\right)^2\right.\right. \\ \nonumber &+&\left.\left.\frac{6(\gamma-1)}{\gamma(\gamma+1)^2}\frac{\Phi''(r_c)}{\Phi'(r_c)}-\frac{6(2\gamma-1)}{\gamma^2(\gamma+1)^2}\right) -  
\Phi''(r_c)+\frac{3\lambda^2}{r_c^4}\right]^{1/2} . 
\end{eqnarray} 
%
%
%

For certain values of $ [\mathcal{E}, \lambda, \gamma, a] $, one obtains more than one (maximum three) physically acceptable ($ r_c > r_+ $) values of the root of the corresponding 

\begin{equation}
\mathcal{E} - f(r_c, \lambda, \gamma) = 0
\end{equation}

One can have integral accretion solutions (stationary) with one critical point (saddle type) or three critical points where one center type critical point is found in between two saddle type critical points. In subsequent sections, we shall develop an eigenvalue scheme to identify which critical point will be of what category. It is to be noted in this context that a physical flow can be constructed only through a saddle type critical point. \\

At this stage the fundamental distinction between a formal multi-critical configuration and a realizable multi-transonic flow requires certain clarification. For stationary axisymmetric accretion, multi-critical flow refers to the orientation for which the algebraic solution of the equation expressing the form of the energy first integral of motion evaluated at the critical point provides three formal roots for $ r_c $ all of them being real positive and located outside $ r_+ $. Such configurations, however, does not necessarily produce multi-transonic flows. Critical point behavior is a formal property of a differential equation of certain class describing certain behaviors of the first order autonomous dynamical systems \cite{js99}. On the other hand, transonicity is a physical property which ensures that the flow makes a smooth continuous transition for one state of sonicity to the other (from sub / super to the super / sub sonic transition). Considering the fact that out of three formal critical points, the middle one being the center type and hence not allowing any transonic solutions to be constructed through it, we explain why the realizable multi-transonic solutions form a sub-class of the formal multi-critical configurations. \\

Subsonic flow starting at the large distance becomes supersonic after crossing the saddle type outer sonic point. Once supersonic, flow cannot access another regular sonic point until it is made subsonic again. Certain additional physical mechanism (a discontinuous shock transition in the present work).  A shock-free multi-critical flow for solution is thus a mathematical construct only, for which the in-going accretion solution remains mono-transonic in practice. The physically realizable multi-transonic configuration is obtained by allowing the flow solution constructed through the outer sonic point with the flow solution passing through the inner sonic point through a steady standing discontinuous shock transition. In what follows, we discuss the criteria of such shock formation in somewhat detail.

\section{\bf Shock formation in axisymmetric inviscid accretion}

In this paper we deal with the angular momentum supported shock transitions. Certain kind of perturbation may produce discontinuities in large scale astrophysical flows. Such discontinuities are supposed to be manifested over a surface when certain dynamical and / or thermodynamic flow variables change discontinuously across such surface, and the corresponding surface is marked as the surface of discontinuity. Based on specific boundary conditions to be satisfied across the surface of discontinuity, such surfaces are classified into various categories - the most important (in astrophysical fluid dynamics) being the shock waves or shocks. For adiabatic flow of Newtonian fluid, the boundary conditions satisfied across the shock may be expressed as in \cite{landau},
\begin{equation}
[[\rho u]] = 0
\end{equation}

\begin{equation}
[[p + \rho u^2]] = 0
\end{equation}

\begin{equation}
[[\dfrac{u^2}{2} + h]] = 0
\end{equation}
 
where $ [[f]] $ denotes the difference of any flow variable $ f $ across the shock, i.e. $ [[f]] = f_1 \sim f_2 $ where $ [f_1, f_2] $ are the boundary values of the quantity $ f $ on the two sides of the shock surface. \\

Supersonic astrophysical flows are prone to shock formation phenomena in general and become subsonic after the shock. The angular momentum supported centrifugal potential barrier acts as a repulsive agent (against the attractive force of the gravity) to break the supersonic incoming flow (matter heading toward a black hole for our case). Such shock formation in accretion astrophysics provides an effective mechanism for conversion of a significant amount of the ordered gravitational potential energy into 'observable' radiation by randomizing the inflow. The issue of shock formation in this work will be addressed along two different directions. The energy preserving Rankine - Hugoniot \cite{rankine,hugoniot1,hugoniot2,landau,salas} shocks will be considered for the adiabatic flow, where the post shock temperature shoots up compared to the position dependent temperature of the pre-shock flow. Higher post-shock temperature puffs up the disc structure which mimics (on local scale) a thick quasi-spherical configuration. The shock thickness is considered to be infinitesimally small - in principle we deal with shocks with shocks with zero effective thickness. \\

For isothermal flow we consider temperature preserving shocks. Flow dissipates energy to maintain the invariance of the temperature. The post-shock flow thickness does not get altered due to such shocks. \\

We define flow variables with the subscript '-' and '+' as the 'pre' and 'post' shock variables, respectively. For constant height flow and flow in conical equilibrium, we obtain the shock condition as

\begin{equation}
\label{shockCO}
\left[ \left[  \dfrac{(\gamma M^2 + 1)^2}{2M^2 + (\gamma-1)M^4} \right]  \right] = 0
\end{equation}
and for flow in vertical equilibrium 

\begin{equation}
\left[ \left[ \dfrac{[M^2 (3 \gamma - 1) + 2]^2}{2M^2 + (\gamma - 1)M^4} \right] \right] = 0
\end{equation}

For isothermal flow, the shock conditions are found to be model independent, i.e., independent of the flow geometry and can be expressed as 

\begin{equation}
\label{shock condition iso 1}
\rho_+ u_+ = \rho_- u_- 
\end{equation}

\begin{equation}
\label{shock condition iso 2}
P_+ + \rho_+ u_+^2 = P_- + \rho_- u_-^2 
\end{equation}

\begin{equation}
\label{shock condition iso 3}
M_+ + \dfrac{1}{M_+} = M_- + \dfrac{1}{M_-}
\end{equation}
\begin{figure}[h!]
\centering
\includegraphics[scale=0.7]{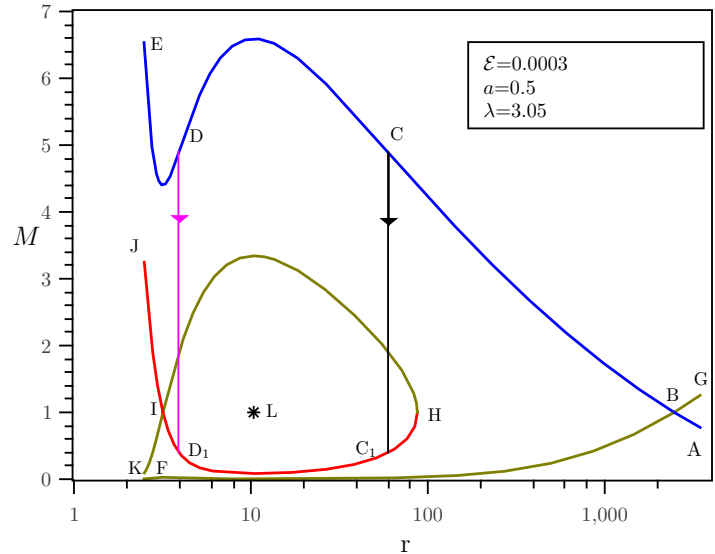} 
\caption{Phase portrait of  polytropic transonic accretion in conical geometry}\label{adiaphs}
\end{figure}

For conical polytropic flow, we show a representative phase portrait for multi-transonic shocked accretion in fig.~\ref{adiaphs} for $ [\mathcal{E} = 0.0003, \lambda = 3.05, a =0.5 ] $. \textbf{ABCDE} is the transonic accretion branch through the outer sonic point \textbf{B}. The solution \textbf{JHIJK} through the inner sonic point \textbf{J} does not connect the outer boundary of  $r$ at infinity. But Rankine-Hugoniot standing shock condition connects two branches of solution at \textbf{CC$_1$}  and \textbf{DD$_1$}. Hence the accretion flow through the outer sonic point may jump discontinuously through \textbf{DD$_1$} to the other branch with higher specific entropy being subsonic again and through that branch \textbf{D$_1$ JI} it may become again supersonic. The other branch \textbf{FBG} in the fig.~\ref{adiaphs} does not have much relevance in the context of accretion flow.  

\begin{figure}[h]
\centering
\includegraphics[scale=0.7]{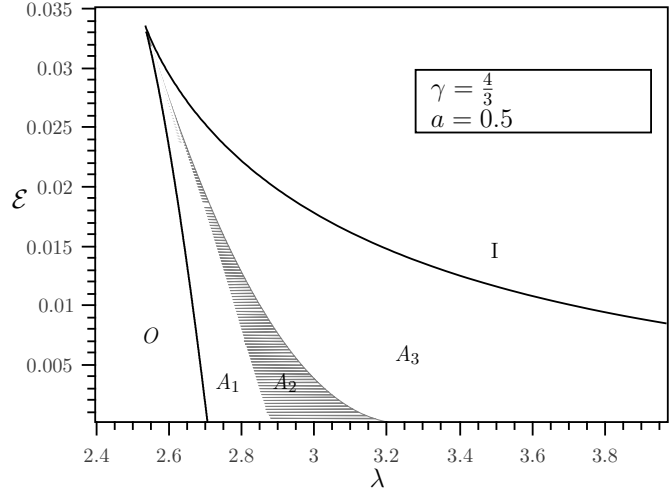} 
\caption{The shaded region in $\mathcal{E}-\lambda$ space allowing shock for  the conical model}\label{shkadia}
\end{figure}

The Rankine - Hugoniot condition can be solved to obtain the region of the generic $ [\mathcal{E}, \lambda, \gamma, a] $ which supports the formation of steady standing shocks. In fig.~\ref{shkadia}, we have shown a two - dimensional projection of the four - dimensional parameter space (the $ [\mathcal{E}, \lambda] $ subspace of the $ [\mathcal{E}, \lambda, \gamma, a] $ space for $ \gamma = \frac{4}{3} $ and $ a = 0.5$) for conical flow. 

Region marked by O and I provide monotransonic solutions exclusively passing through the outer type (located far away from the black hole) and the inner type (located close to the black hole) sonic points, respectively. Region $ A_1 + A_2 $ represents the multi-critical solutions for which the condition $ \dot{\mathcal{M}}_{\text{in}} > \dot{\mathcal{M}}_{\text{out}} $ is satisfied. $ [\mathcal{E}, \lambda, \gamma, a]_{A_2} $ represents the true multi-transonic region which represents the shock formation. $ [\mathcal{E}, \lambda, \gamma, a]_{A_3} $ provides multi-critical flow with $ \dot{\mathcal{M}}_{\text{out}} > \dot{\mathcal{M}}_{\text{in}} $. In such region, however, accretion flow remains mono-transonic. \\

\begin{figure}[h]
\centering
\begin{subfigure}
 \centering
\includegraphics[scale=0.55]{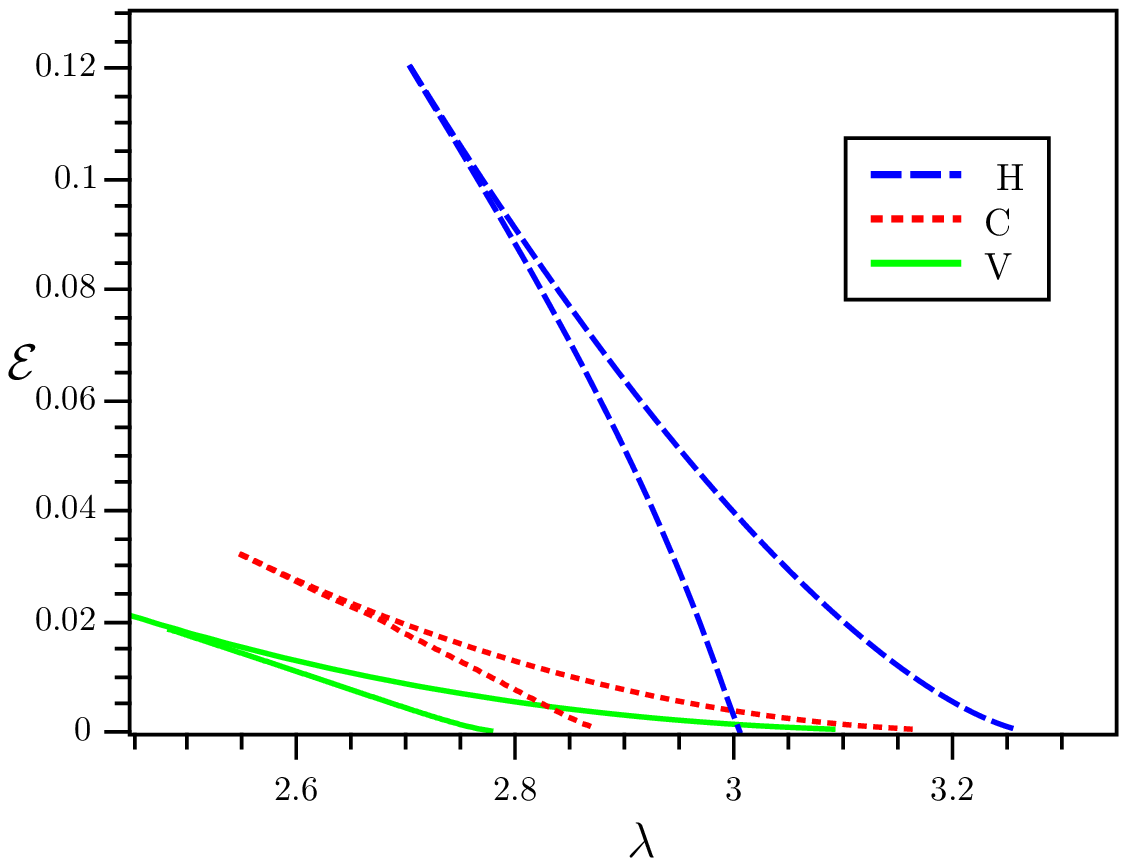} 
\label{modadiashk}
\end{subfigure}
\begin{subfigure}
\centering
\includegraphics[scale=0.55]{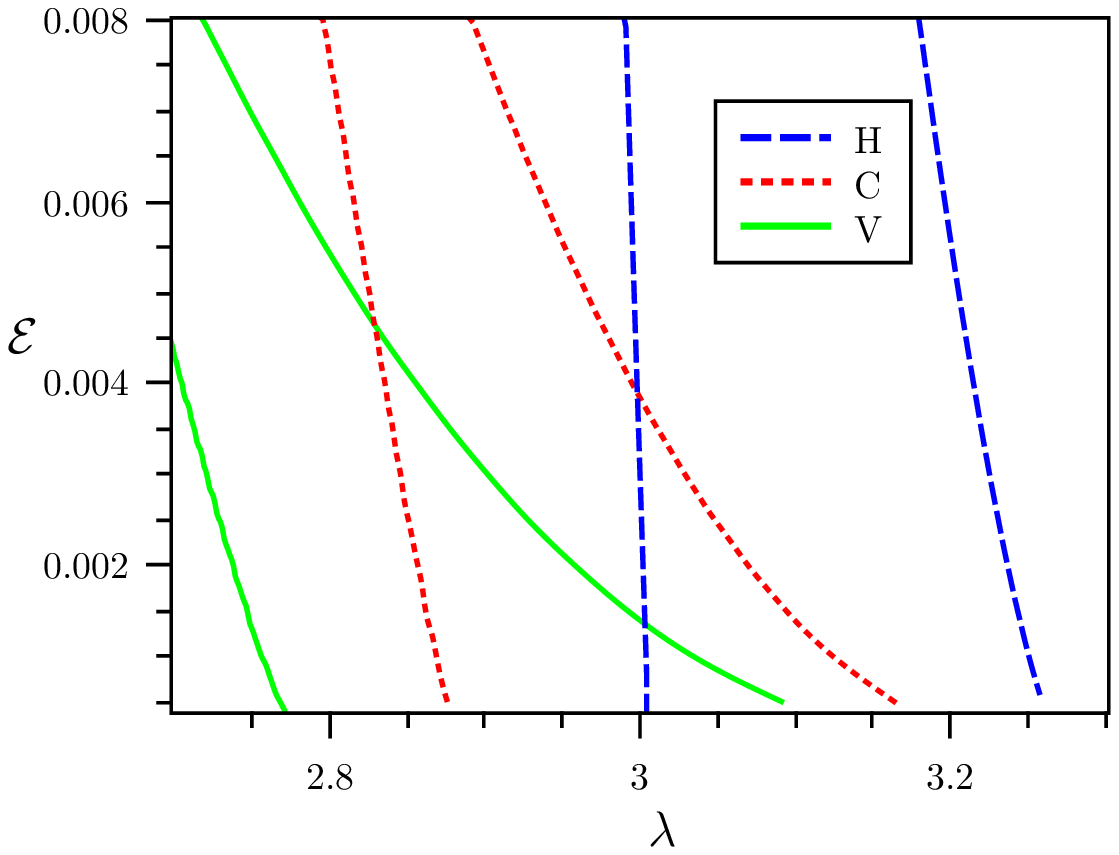} 
\label{mgshkadia}
\end{subfigure}
\caption{Figure in the left side compares the shock allowing regions for three models. Dashed (blue in online version) lines for constant height, dotted  (red in online) lines for conical and solid (green in online) lines for vertical equilibrium model. The figure in right side magnifies the common region}
\end{figure}

In fig.~\ref{modadiashk}, we plot the  $ [\mathcal{E}, \lambda, \gamma, a]_{shock} $ for three different flow models as shown in the figure. 

\begin{figure}[htb]
\centering
\includegraphics[scale=0.6]{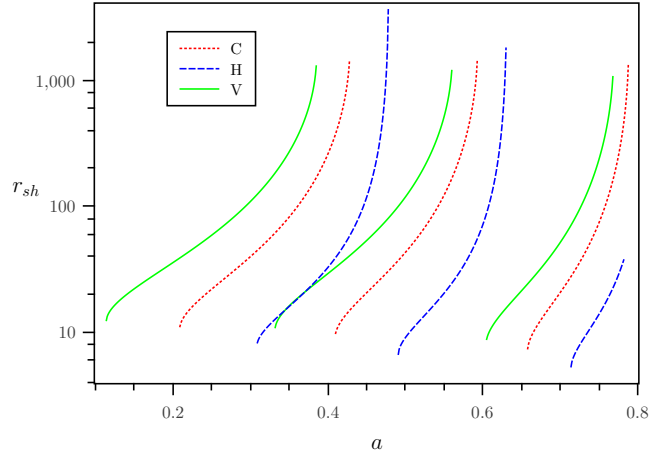} 
\caption{Variation of shock location with spin for three models at different values of 1 $\lambda$ (see text). Line types and color scheme are same as previous figure}\label{shklocdia}
\end{figure}
We find that the constant height flow accommodates standing shock for a relatively higher value of $ \lambda $ ( $ \lambda $ still remains sub-Keplerian, though). Fig.~\ref{mgshkadia} shows the region of  $ [\mathcal{E}, \lambda, \gamma, a]$  common to all three flow geometries which allows the shock formation. 

\begin{figure}[h!]
\centering
\includegraphics[scale=0.6]{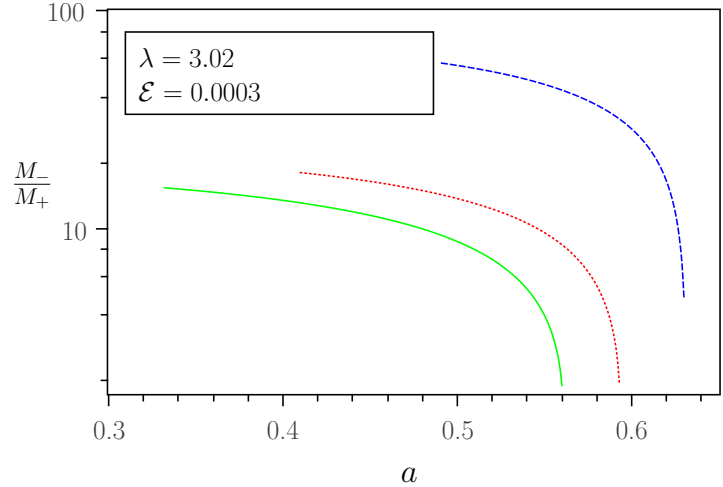} 
\caption{ Variation of shock strengths with spin for three models (Line types and color scheme same).}\label{shkstadia}
\end{figure}
This mutually overlapping shock forming  $ [\mathcal{E}, \lambda] $  region, and its generalized version in the four dimensional  $ [\mathcal{E}, \lambda, \gamma, a] $ space is very crucial finding of our present work. If one requires comparison between three different models based on any physical process / value of the flow variable corresponding to the multi-transonic shocked accretion, one has to take  $ [\mathcal{E}, \lambda] $  from the subspace as shown in fig.~\ref{mgshkadia}. In fig.~\ref{shklocdia}, we show the dependence of the shock location on the black hole spin parameter for constant height flow (hereafter any variation in the figure corresponding to the constant height flow will be drawn using blue color and will be marked by $ H $. For conical flow and flow in hydrostatic equilibrium similar variations will be depicted using red and green color, respectively, and will be marked by $ C $ and $ V $, respectively), conical flow and flow in vertical equilibrium for $ [\mathcal{E} = 0.0003, \gamma = 4/3] $ and for three sets of specific angular momentum, $ \lambda = 2.6 $ (left most set of curves), $ \lambda = 3.02 $ and $ \lambda = 3.3 $ (right most set of curves), respectively. 
\begin{figure*}[h!]
\centering
\begin{subfigure}
 \centering
\includegraphics[scale=0.55]{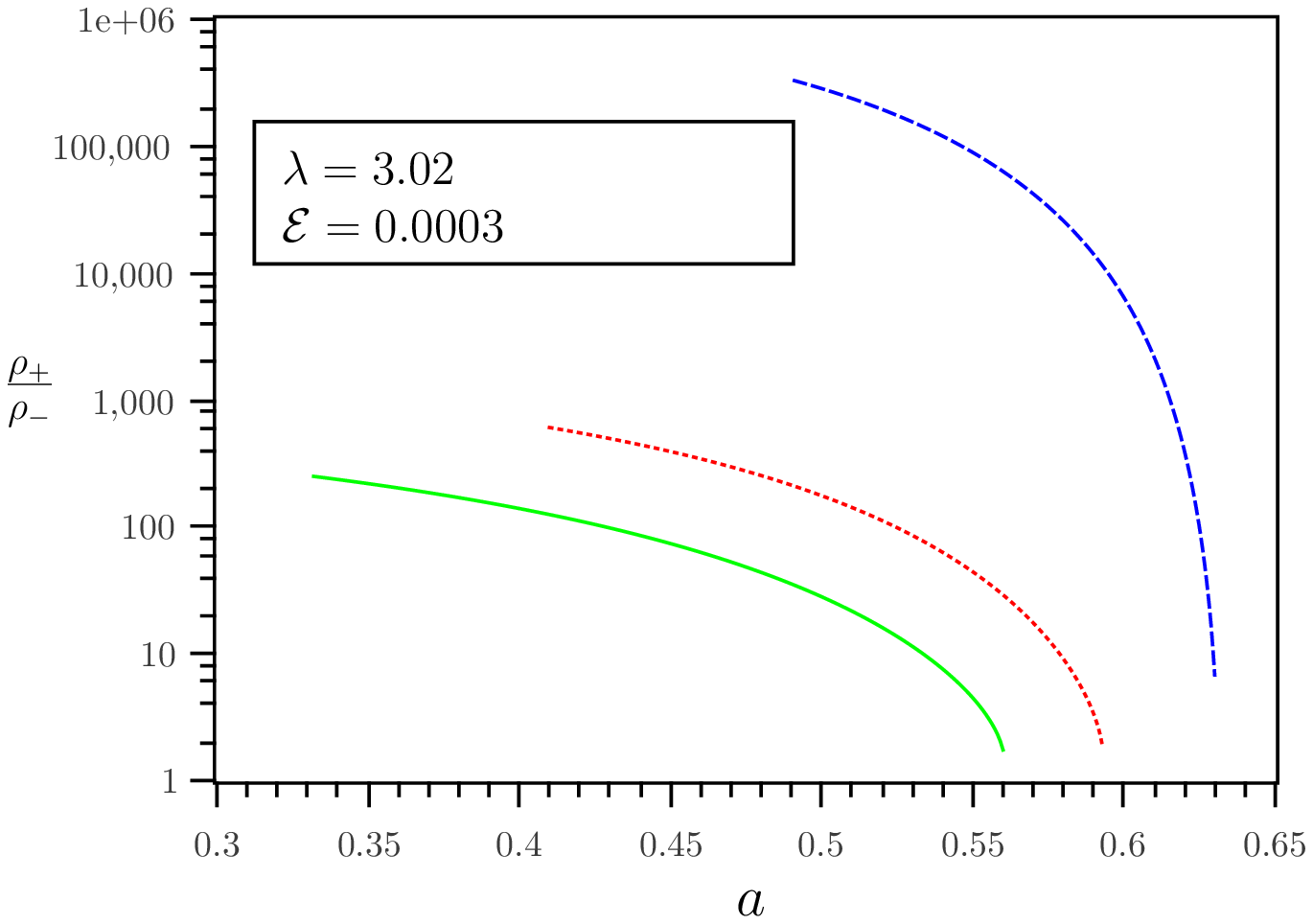} 
\end{subfigure}
\begin{subfigure}
\centering
\includegraphics[scale=0.5]{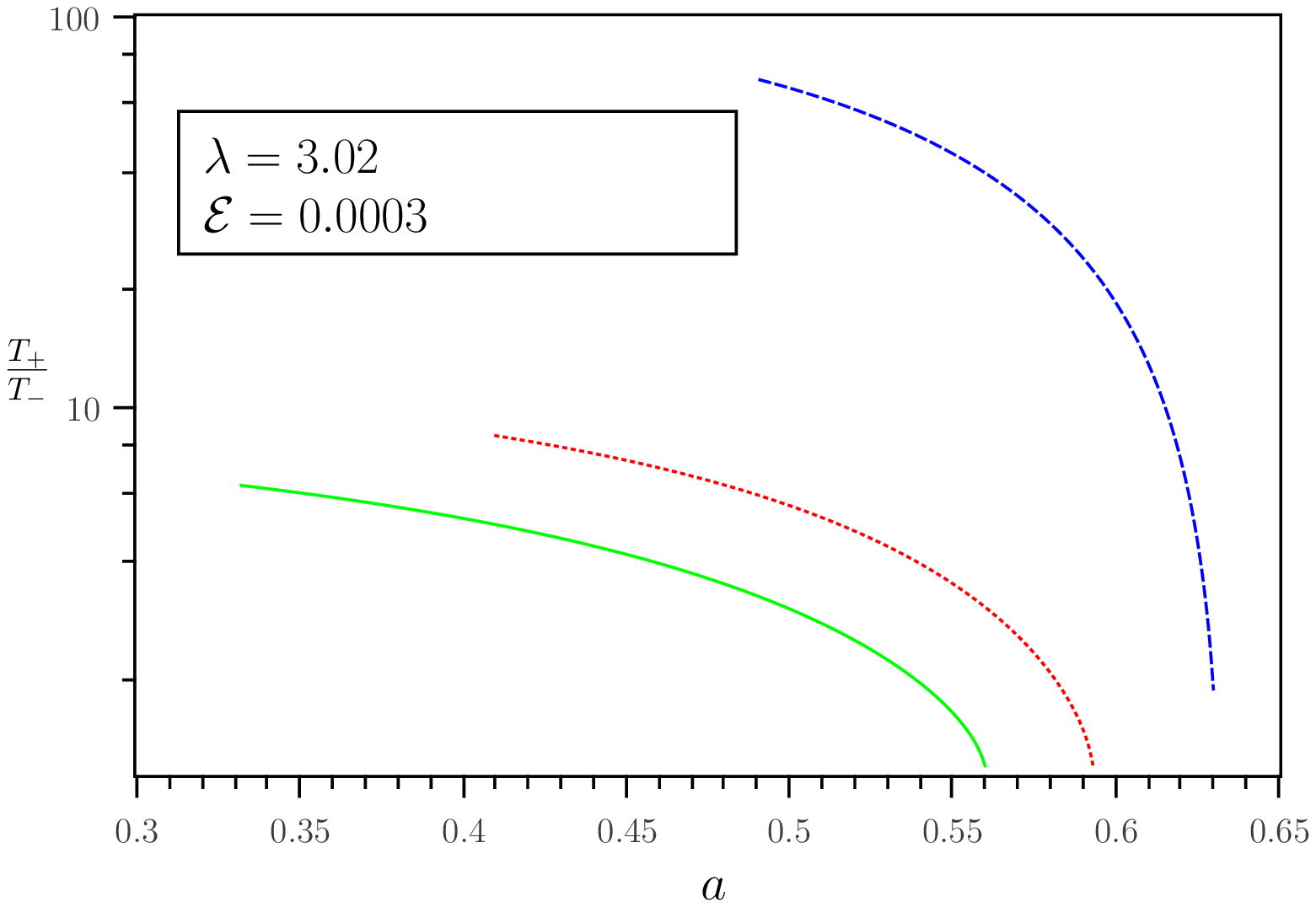} 
\end{subfigure}
\begin{subfigure}
\centering
\includegraphics[scale=0.5]{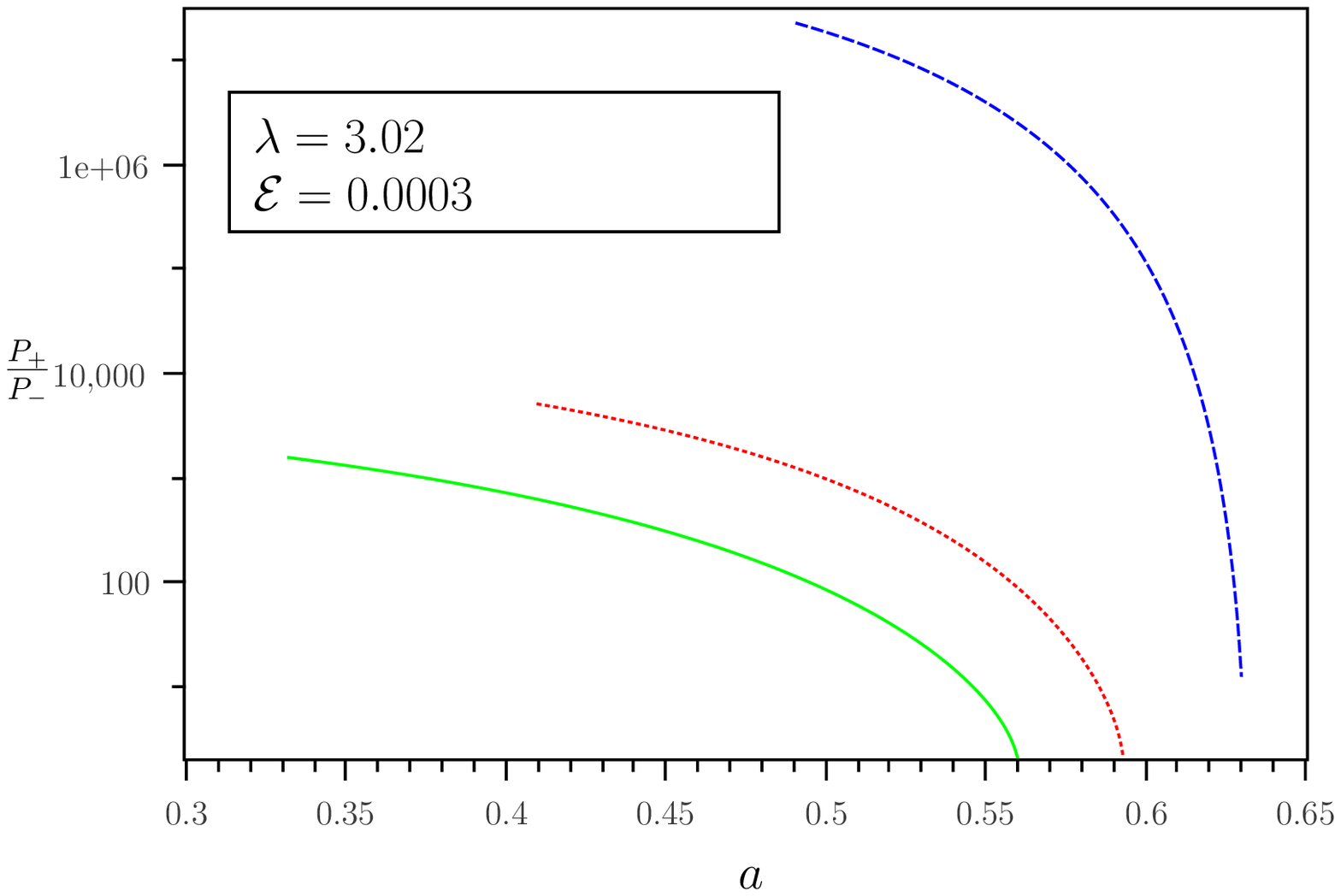} 
\end{subfigure}
\caption{Variation of different shock parameters with $a$, in comparing three models (line types and color scheme same)}\label{shkprmadia}
\end{figure*}

We find that the shock location non-linearly correlates with black hole spin (same result has been obtained for full general relativistic flow in the Kerr metric, see, e.g., Figure ~1 in \cite{Das2014}. \\

It is intuitively obvious that the shocks located closer to the black hole event horizon is stronger. This is because of the fact that more amount of gravitational potential energy is available to be released for smaller shock location. We quantitatively demonstrate such finding through fig.~\ref{shkstadia}, where we have plotted the shock strength (defined by the pre to post shock values of the Mach number) as a function of the Kerr parameter.

Clearly, stronger shocks are produced for smaller shock locations. It is also observed that for the same shock location, the shock is strongest for constant height flow and is weakest for flow in vertical equilibrium. In fig.~\ref{shkprmadia}, the variation of the ratio of post to pre-shock quantities have been shown as a function of the black hole spin.

The density compression ratio $ \rho_+/\rho_- $, ratio of temperature $ T_+/T_- $ and pressure $ p_+/p_- $ anti-correlate with the Kerr parameter for obvious reason (that stronger, hotter and intensely compressed post shock flows are generated closer to the event horizon).

\section{\bf Isothermal Flow}
\begin{figure}[h!]
\centering
\includegraphics[scale=0.6]{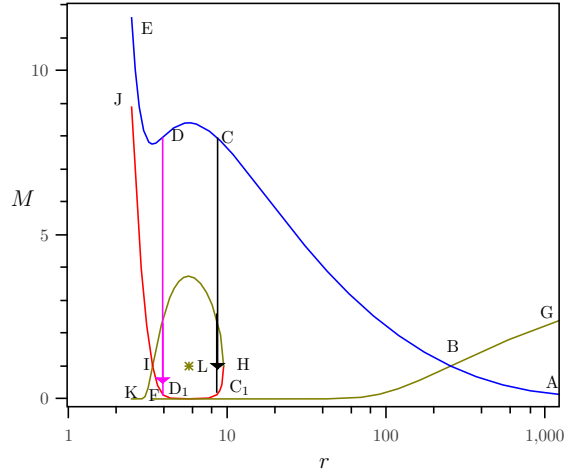}
\caption{Phase plot for isothermal accretion with isothermal shocks, similar to polytropic one (see fig.~\ref{adiaphs})}\label{isophs}
\end{figure}
A typical shocked multi-transonic flow topology is shown in fig.~\ref{isophs} for $ [T_{10} =1 , \lambda = 3.05 , a = 0.5] $ .

\begin{figure}[h!]
\centering
\includegraphics[scale=0.7]{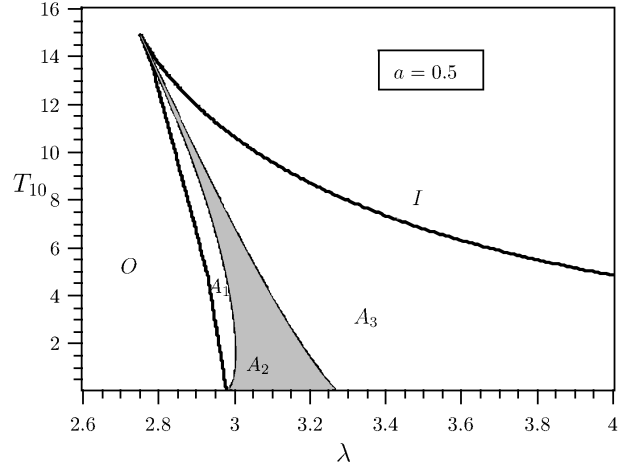}
\caption{Region of parameter space allowing isothermal shock (see text)}\label{shkiso}
\end{figure}
The two dimensional projection of the three dimensional $ [T, \lambda, a] $ parameter space is shown in fig.~\ref{shkiso} for conical model with the black hole spin parameter $ a = 0.5 $.

\begin{figure}[hbtp]
\centering
\begin{subfigure}
\centering
\includegraphics[scale=0.55]{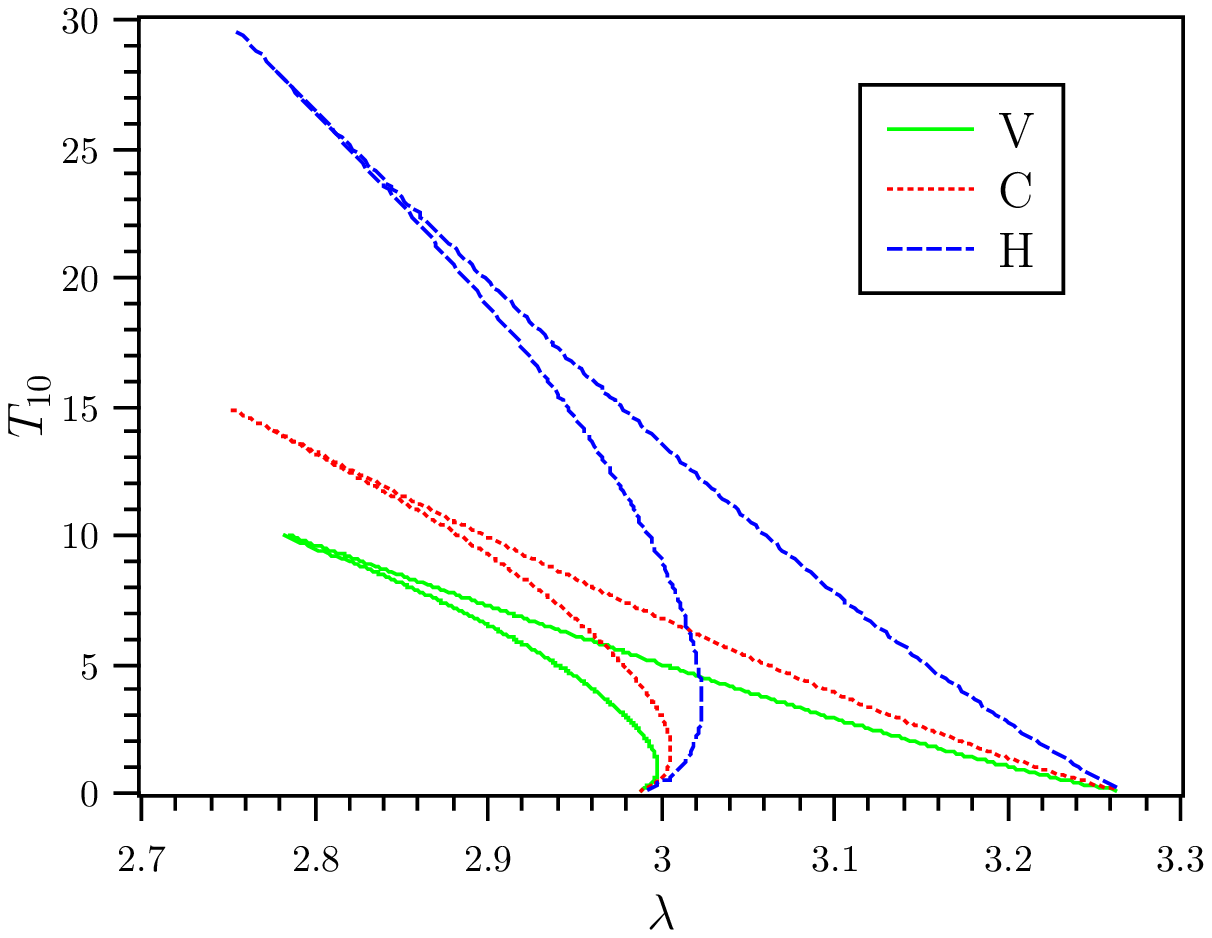}
\end{subfigure}
\begin{subfigure}
\centering
\includegraphics[scale=0.55]{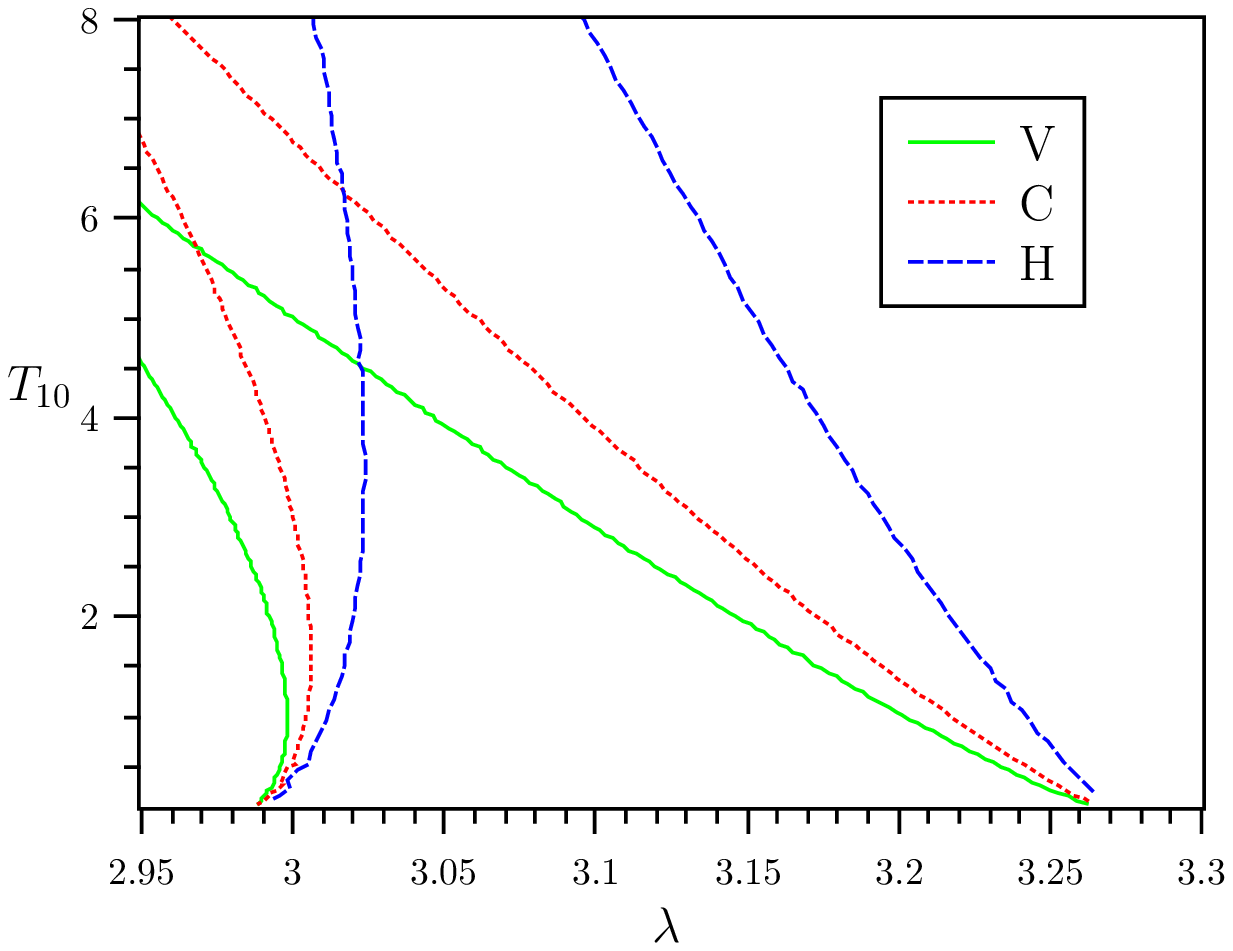}
\end{subfigure}
\caption{Comparison of regions allowing isothermal shock in parameter space for different models (color scheme same). The figure in right magnifies the common region of the figure in left.}\label{shkmodiso}
\end{figure}

The symbols O and I represent mono-transonic accretion constructed through the outer and the inner sonic points, respectively. $ (A_1+A_2) $ represents the multi-critical region for which $ \xi_{in} > \xi_{out} $ and $ A_3 $ represent the multi-critical region for $ \xi_{in} < \xi_{out} $. For $ A_1 $ and $ A_3 $ , however, ingoing accretion flow is mono-transonic, passing through the outer sonic point for $ [T, \lambda]_{A_1} $ and through the inner sonic point for $ [T, \lambda]_{A_3} $. $ [T, \lambda]_{A_2} $ provides the true multi-transonic accretion endowed with a temperature preserving standing shock. fig.~\ref{shkmodiso} shows the parameter space spanned by $ T $  and $ \lambda $ for which the shock formation is possible for three different flow geometries. 

Once again, the constant height flow allows a larger $ [T, \lambda] $ region for the shock formation compared to other two flow geometries.  We have a large overlapping region here for the formation of the isothermal shock. The enlarged (sufficiently zoomed in) common region is shown in fig.~\ref{shkmodiso}.
\begin{figure}[hbtp]
\centering
\includegraphics[scale=0.6]{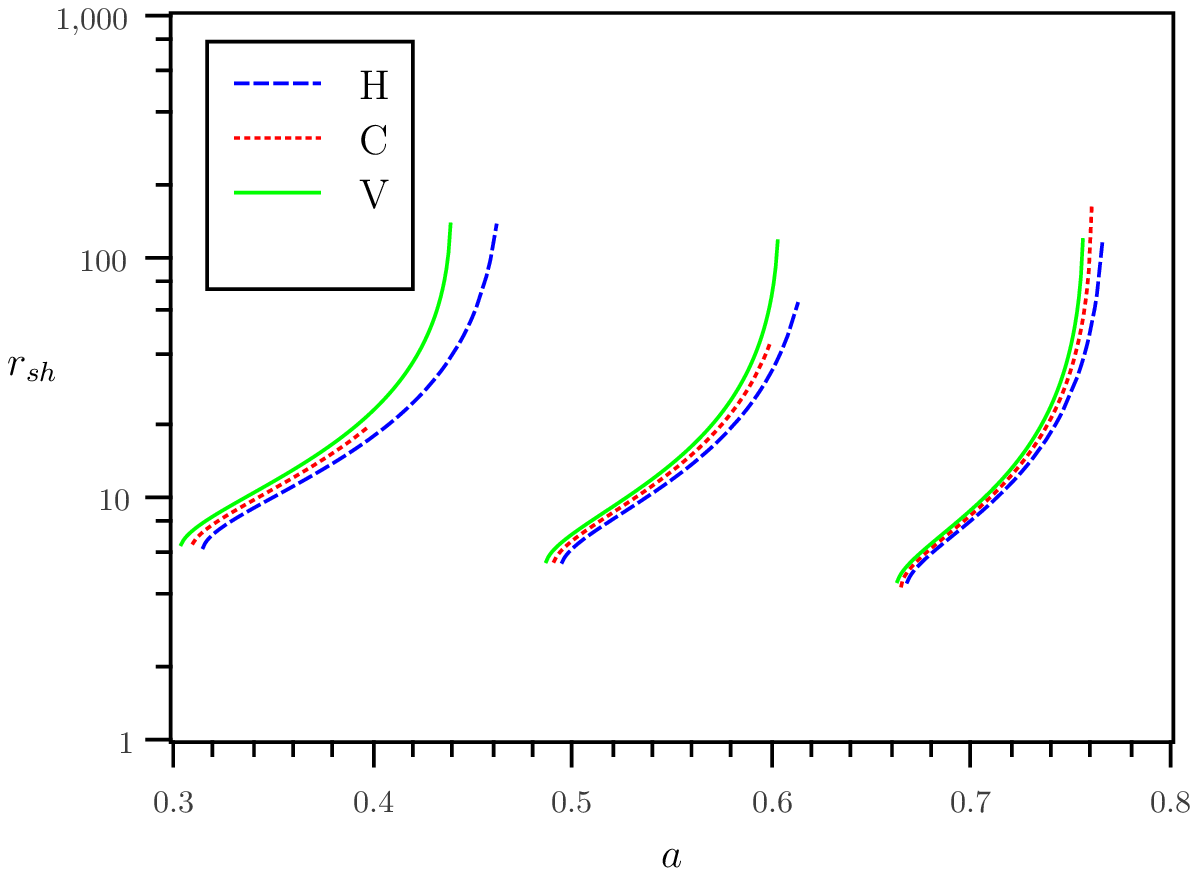}
\caption{Variation of shock location with $a$ for three models.}\label{shklociso}
\end{figure}
\begin{figure}[hbtp]
\centering
\includegraphics[scale=0.6]{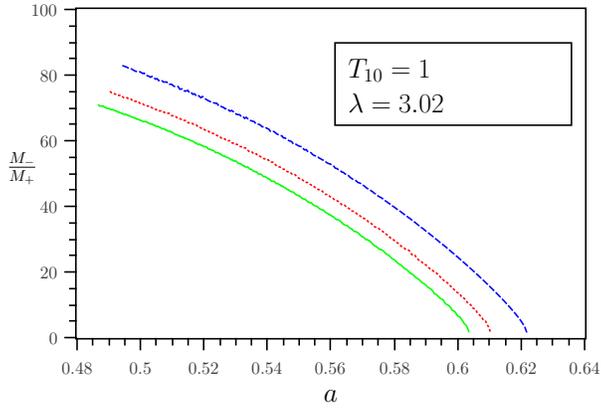}
\caption{Variation of shock strength with $a$ for three models.}\label{shkstiso}
\end{figure}

 The variation of the shock location with the black hole spin parameter is shown in fig.~\ref{shklociso}.

 The $ r_{\text{sh}} - a $ variation follows the same trend as obtained in the polytropic flow. 
The variation of the shock strength with the Kerr parameter has been shown in fig.\ref{shkstiso}.

\begin{figure}[hbtp]
\centering
\begin{subfigure}
\centering
\includegraphics[scale=0.5]{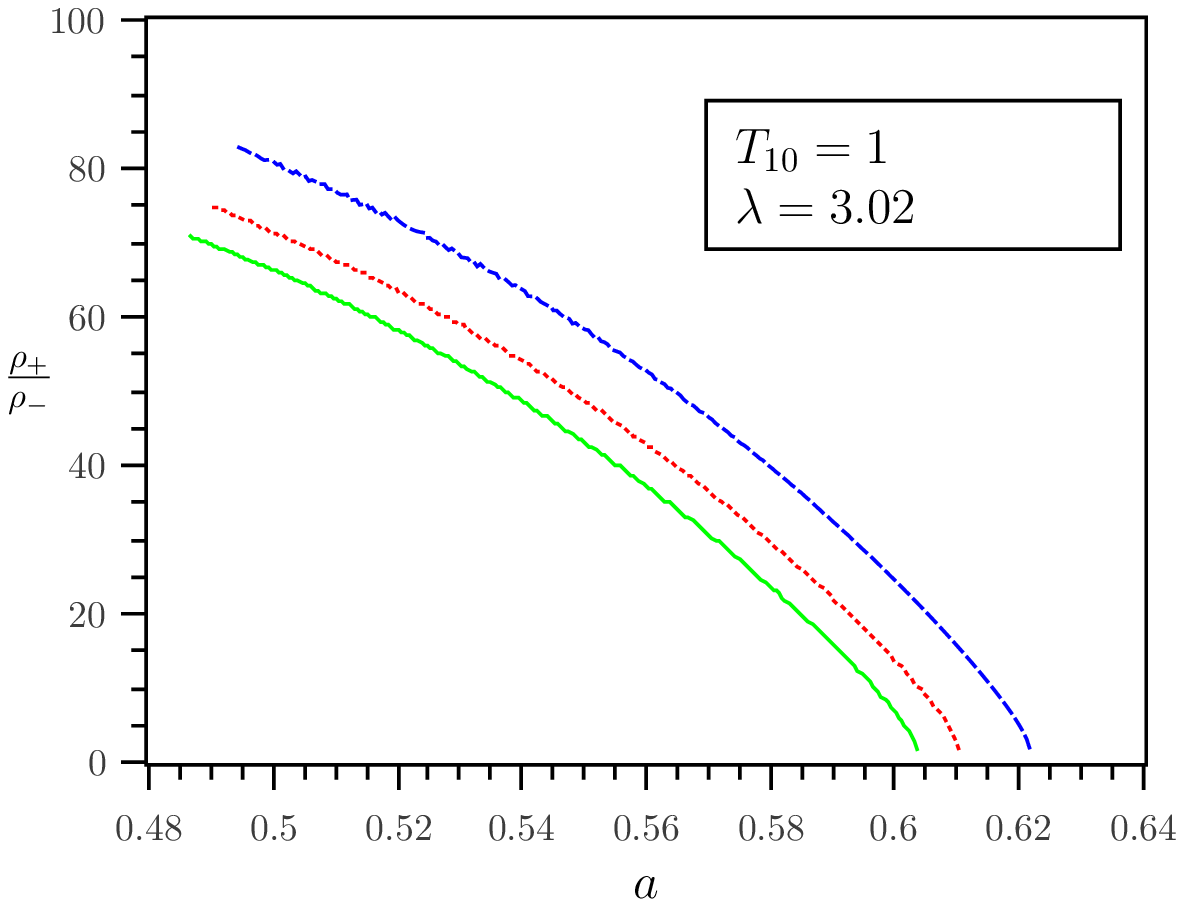}
\end{subfigure}
\begin{subfigure}
\centering
\includegraphics[scale=0.55]{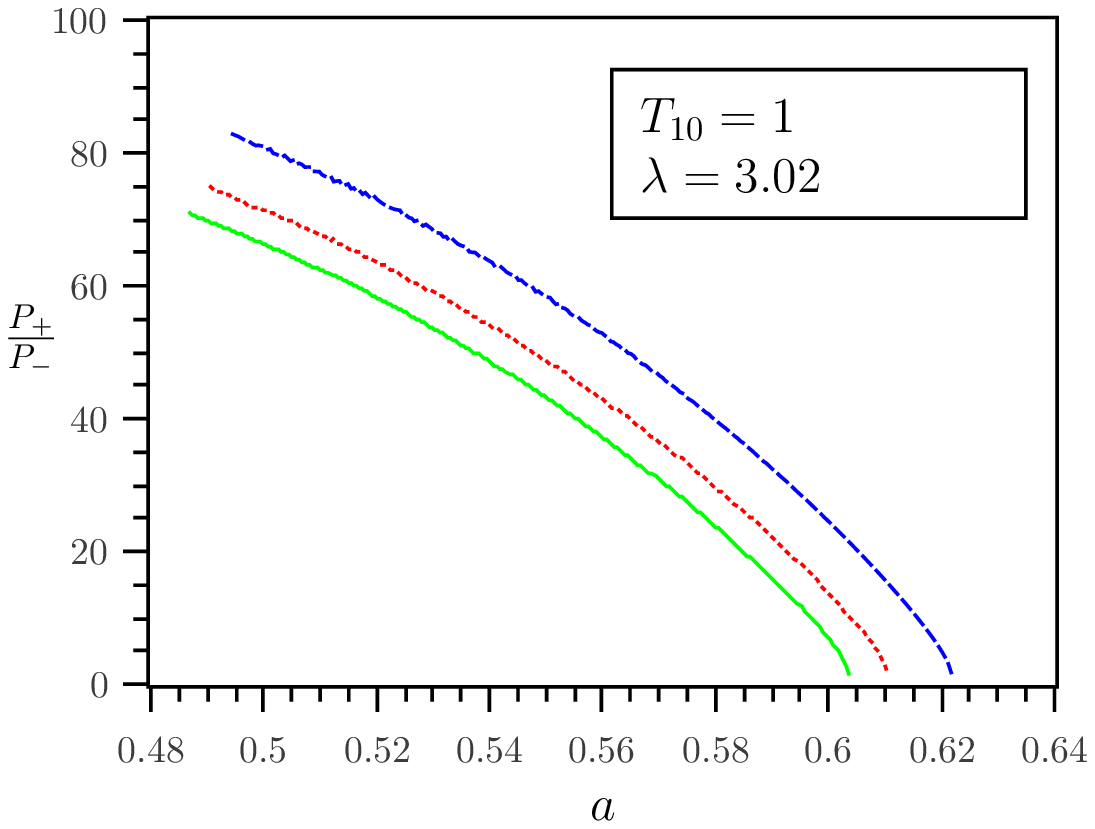}
\end{subfigure}
\caption{Variation of shock parameters with $a$ for three models.}\label{shkprmiso}
\end{figure}
Stronger shocks are formed closer to the horizon and constant height flow produces the strongest possible shock variation of the density compression ratio as well as the ratio of the post to pre-shock pressure on the black hole spin is shown in fig.~\ref{shkprmiso}.

\section{\bf Stability of the stationary flow}
The stability of the stationary flow configuration is to be ensured for the validity of this model. Following the method depicted in refs.\cite{Petterson1980,Theuns1992,ray03a,crd06,nard12}, one introduce time dependent perturbation of the form $u(r,t) = u_0(r) + \tilde{u}(r,t)$ and 
$\rho (r,t) = \rho_0 (r) + \tilde{\rho}(r,t)$ where the quantities with suffix $_0$ correspond to the values for stationary flow at a particular $r$ and the the additional time dependent parts under tilde are the corresponding perturbing parts. These quantities are introduced into the time dependent counterparts of continuity equation, eq.~\eqref{equation of continuity} and Euler equation, eq.~\eqref{euler}. Following the same line of argument as mentioned in the ref.[] with general form of pseudo-potential $\phi$, one may show that within the time evolution equation of the perturbing part of the mass accretion rate $f=\rho urH$, the potential  $\phi$ does not enter, as long as it being a function of radial distance $r$ only. Thus the time evolution equation of the perturbing part  $\tilde{f}$ becomes ref.~\cite{nard12} \begin{equation}
\prt_\mu \left( {\mathrm{f}}^{\mu \nu} \prt_\nu 
\tilde{f}\right) = 0 \,,
\end{equation}
in which the Greek indices are made to run from $0$ to $1$, with 
the identification that $0$ stands for $t$, and $1$ stands for $r$; and \begin{equation}
\label{matrix}
\mathrm{f}^{\mu\nu}=\frac{u_0}{f_0}\left(\begin{array}{cc}
1 & u_0 \\ 
u_0 & u_0^2-\sigma c_{s0}^2
 \end{array} \right);
\end{equation}
$\sigma$ being a model dependent parameter such that $\sigma=1$ for CH and CM while $\sigma=2/(\gamma+1)$ for VE. Hence following the argument of refs.\cite{crd06,nard12}, basing upon the identical time evolution equation, it may be concluded that the stationary flow is stable in this case too.

Moreover comparing it with wave equation of a scalar (say $\psi$) in curved space time i.e., \begin{equation}
\frac{1}{\sqrt{-g}}\left[\prt_{\mu}\left(\sqrt{-g}g^{\mu\nu}\prt_{\nu}\right)\psi\right]=0
\end{equation} (where $g=\mathrm{det}(g_{\mu\nu})$) following  [Visser1998], one may form an effective metric for the acoustic perturbation (henceforth to be called as acoustic metric, $G_{\mu\nu}$) as $\sqrt{-\mathrm{det}(G_{\mu\nu})}G^{\mu\nu}\equiv\mathrm{f}^{\mu\nu}$.

\section{\bf Acoustic Surface Gravity : Dependence on the black hole spin parameter}

The metric elements $ G_{\mu \nu} $ corresponding to the sonic geometry is obtained as a consequence of the linear stability analysis. This we have demonstrated in the previous section. The acoustic surface gravity $ \kappa $ for the stationary background fluid can be obtained as (see ref.\cite{Bilic2014})
\begin{equation}
\kappa =\left\vert\sqrt{(1+2\phi(r))\left(1-\frac{\lambda^{2}}{r^2}-2\phi(r)\frac{\lambda^{2}}{r^2}\right)}\left(\frac{1}{1-{c_s}_c ^2}\left[\left.\frac{du}{dr}\right|_c-\left.\frac{dc_s}{dr}\right|_c\right]\right)\right\vert. 
\end{equation}

\begin{figure}[h!]
\centering
\includegraphics[scale=0.6]{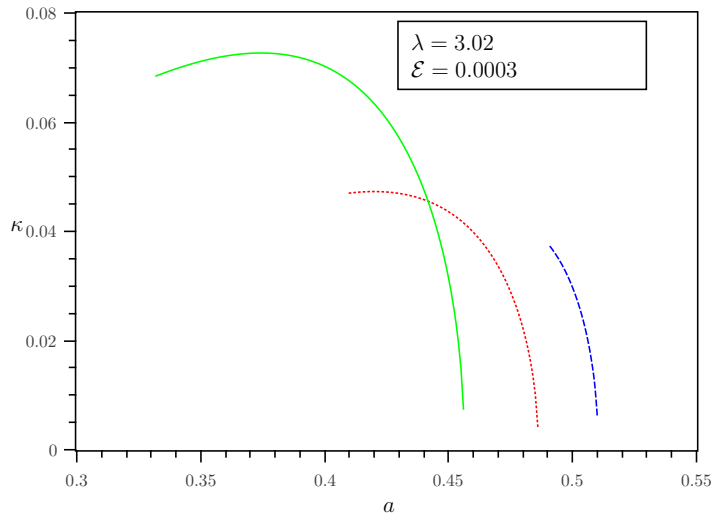}
\caption{Variation of surface gravity with $a$ for three polytropic flow models.}\label{kapadi}
\end{figure}
As is obvious from the explicit expression of $\kappa$ , the surface gravity is a function of our four parameter initial boundary condition governing the flow, i.e., 
$ \kappa = \kappa[\mathcal{E}, \lambda, \gamma, a] $. \\ 
For a fixed set of values of $ [\mathcal{E}, \lambda, \gamma] $, we would like to calculate $\kappa$ at the inner sonic point for a multi-transonic flow for a range of Kerr parameters to study the dependence of the acoustic surface gravity on black hole spin. $ \kappa - a $ relationship could also be demonstrated for the outer acoustic horizon for a shocked flow. However, $\kappa$ at the outer acoustic horizon has numerical value way less compared to that of calculated at the inner acoustic horizon. This is a generic property (that $ \kappa_{\text{in}} >> \kappa_{\text{out}} $) found independent of the nature of the background space-time metric, geometric configuration of the accretion flow, as well as the thermodynamic equation of state used to describe the matter flow in general. This indicates that the numerical value of the acoustic surface gravity correlates with the strength of the gravitational attraction of the background gravitational field. Also to mention in this context that $ \kappa_{out} $ is not sensitive enough on 'a' when evaluated at the outer acoustic horizon. This is intuitively obvious because at the outer acoustic horizon (which forms a large distance away from the black hole), space-time becomes asymptotically flat and the effect of the black hole spin does not really affect the dynamics of the flow, and hence the nature of the sonic geometry embedded within. \\

\begin{figure}[hbtp]
\centering
\includegraphics[scale=0.65]{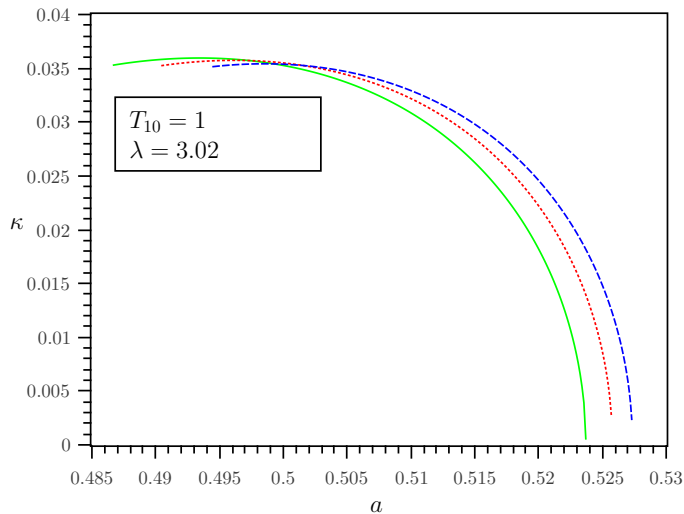}
\caption{Variation of surface gravity with $a$ for three isothermal flow models.}\label{kapiso}
\end{figure}

In fig.~\ref{kapadi} and fig.~\ref{kapiso}, we show the $ \kappa - a $ dependence for adiabatic as well for the isothermal accretion, respectively, for three different flow geometries considered in our work.

Note that the $ \mathcal{E} - \lambda $ parameter space responsible for shock formation allows a rather narrow range of the black hole spin when we consider the multi-transonic parameter space common to all three flow models. Hence for a fixed value of $ [\mathcal{E}, \lambda, \gamma] $ or $ [T, \lambda] $ , to find an appreciable span of 'a' is rather non-trivial while comparing the nature of $ \kappa - a $ variation for different flow configurations.
One can, however, bypass such constraints by studying the dependence of the acoustic surface gravity as a function of the black hole spin for mono-transonic flow constructed through the inner acoustic horizon. For such flow configuration, the entire range of 'a', i.e., $ [-1, 1] $, may be accessed and hence a more comprehensive information about $ \kappa - a $ variation may be obtained.  

\section{\bf Acknowledgement}
S. Sen would like to acknowledge the kind hospitality
provided by HRI, Allahabad, India, under a visiting student
research programme. The visits of S. Sen, SR and SN at HRI was partially supported
by astrophysics project under the XIIth plan at HRI.  TKD would like to acknowledge the hospitality provided by Department of Physics, Sarojini Naidu College for Women, Kolkata.
The work of SN was partially supported by UGC MRP No.PSW-163/13-14.

\end{document}